\documentclass{svproc}
\usepackage[hyphens,spaces,obeyspaces]{url}

\usepackage{hyperref}

\usepackage{booktabs} 
\usepackage{multirow}
\usepackage{graphicx}
\usepackage{diagbox} 
\usepackage{enumitem}
\usepackage{breakcites}
\usepackage{color}
\usepackage{amsmath}
\usepackage[ruled, vlined, longend]{algorithm2e}

\begin{document}
\mainmatter              
\title{Simplifying Formal Proof-Generating Models with ChatGPT and Basic Searching Techniques}
\titlerunning{Simplifying Proof Generation with ChatGPT and Basic Search}  
%
\author{Sangjun Han\inst{1} \and Taeil Hur\inst{2} \and Youngmi Hur\inst{1}\inst{,3} \and Kathy Sangkyung Lee\inst{1} \and Myungyoon Lee\inst{1} \and Hyojae Lim\inst{4}}

\authorrunning{S. Han et al.}

\institute{Yonsei University, Seoul, South Korea\\
\email{\{qkqhwl4,yhur,kathy,yooni7272\}@yonsei.ac.kr}
\and
JENTI Inc., Seoul, South Korea\\
\email{taeil.hur@jenti.ai}
\and
Korea Institute for Advanced Study, Seoul, South Korea
\and
Johann Radon Institute for Computational and Applied Mathematics, Linz, Austria\\
\email{hyojae.lim@oeaw.ac.at}}

\maketitle              

\begin{abstract}
The challenge of formal proof generation has a rich history, but with modern techniques, we may finally be at the stage of making actual progress in real-life mathematical problems. This paper explores the integration of ChatGPT and basic searching techniques to simplify generating formal proofs, with a particular focus on the miniF2F dataset. We demonstrate how combining a large language model like ChatGPT with a formal language such as Lean, which has the added advantage of being verifiable, enhances the efficiency and accessibility of formal proof generation. Despite its simplicity, our best-performing Lean-based model surpasses all known benchmarks with a 31.15\% pass rate. We extend our experiments to include other datasets and employ alternative language models, showcasing our models' comparable performance in diverse settings and allowing for a more nuanced analysis of our results. Our findings offer insights into AI-assisted formal proof generation, suggesting a promising direction for future research in formal mathematical proof.

\end{abstract}

\section{Introduction}
Writing mathematical proofs in formal language has been a research topic for many years. Notably, we recall McCarthy's symbolic computation on machines \cite{LISP}, Hoare's axiomatic approach to computer programming~\cite{Hoare}, and DeMillo, Lipton, and Perlis's skepticism towards formal proofs of theorems \cite{DeMillo79}.
Due to the surprising performance of AI algorithms witnessed over the last decade, such as transformers in natural language processing, we believe it is an opportune time to revisit the formal proof challenge with the modern techniques available. As evidenced by our literature review below, we are not alone in this pursuit. 

Our primary interest lies in writing ordinary, workaday mathematical proofs in formal language and, conversely, in using these formal proofs to enhance our everyday mathematical reasoning. With this ultimate goal in mind, we have made progress, albeit slowly, in understanding this issue over the past few years. In this paper, we aim to share some of the insights and experiences we have gained, hoping to contribute to the broader research community's understanding of future directions to explore in tackling this complex problem.

We utilize ChatGPT (GPT-4 and GPT-4 Turbo) together with Lean for a set of experiments, primarily focusing on the miniF2F dataset \cite{minif2f}, a standard dataset in formal proof research. Our models achieve surprisingly competent performance for their simplicity, notably our best-performing model, which surpasses most known benchmarks with a 31.15\% pass rate. We evaluate our approach on other datasets to ensure similar effectiveness across various mathematical domains and previously unpublished problems. We also study our search algorithms under another language model, Llemma, displaying the versatility of our approach. 

The outline of the rest of this paper is as follows. The rest of Section 1 broadly discusses the previous work related to our research. In Section~2, we cover some preliminary background material that would be helpful in understanding our work, which includes Lean, miniF2F, and proof search algorithms. We explain the characteristics of our models in Section 3, and Section 4 details our experiments.
We analyze the results of our models with specific examples in Section~5, and present our conclusion and overview in Section~6.

\subsection{Related Work}

Mathematical theorem proving in machine learning has garnered significant attention, particularly for developing artificial intelligence capable of logical reasoning. As a result, various studies have been conducted, with those most closely related to our work outlined below.

\subsubsection{Formal and informal languages}

Languages used for the integration of machine learning with mathematical reasoning can be broadly categorized as either formal or informal.
Informal language has a format that is familiar to the human eye and is thus easily understood and used. 
The authors in \cite{math_machine_intution} propose a machine learning-based framework and leverage informal language to offer mathematical insight, and the models in \cite{tora, NaturalLanguage} are fine-tuned to prove mathematical theorems in informal language. 
However, informal language proofs have the disadvantage of being difficult to verify. To address this issue, many studies have utilized verifiable programming languages, spanning across formal languages such as Lean \cite{ProofNet,pact,curriculum,leandojo}, Metamath \cite{GPTf}, HOL4 \cite{TacticZero}, Isabelle \cite{thor,DraftSketchProve,lego}, Coq \cite{CoqGym}, Python \cite{codex}, or a mix of these environments \cite{HyperTree,copra}. 
These formal languages can represent mathematical ideas precisely, making them easier to utilize in a computing system. However, formal languages differ vastly from the mathematical arguments written by humans, making it difficult to interpret without ample understanding of the formal setting used. Given that we consider the verifiability of proofs to be essential, we choose to work with formal language to construct the models in our study, despite its lower accessibility.

\subsubsection{Large language models with theorem proving}
Large language models (LLMs) have become increasingly useful in natural language processing \cite{gpt4,gpt3,codex,palm,llama}, inspiring many studies on these models.
The researchers from \cite{MIT} use Codex to find solutions for informal language problems by converting the problems into programming tasks, while the authors of \cite{copra} improve their model's performance by composing effective prompts for ChatGPT.
Many also create new models based on LLMs, such as Llemma \cite{llemma}, which achieves high performance in mathematical proof generation by fine-tuning LLaMA \cite{llama} on newly proposed data.
Others, as those in \cite{DraftSketchProve,lego}, link together LLMs and theorem proving by writing formal proofs informed by the informal proofs generated for a mathematical problem.
In this paper, we aim to construct a simple and easily reproducible model that can be tried by anyone.
To this end, we utilize ChatGPT and enhance its performance by modifying existing proof search algorithms specifically tailored for this LLM.

\subsubsection{Development of proof search algorithms}
There have been various efforts towards developing formal language models. Among these, there are many models that have modified the proof search algorithm in innovative ways, often displaying strong performances. A pioneer in this endeavor is GPT-f \cite{GPTf}, which introduced the proof search process and evaluation criteria for the field. 
Some models rely on extra training, such as expert iteration in \cite{curriculum} to tackle problems of increasing difficulty and control the lengths of formal proofs, or reinforcement learning in \cite{TacticZero}.
Other models focus on premise selection to call relevant theorems or other mathematical statements. For example, Thor \cite{thor} utilizes Hammer to aid the proof search process by choosing appropriate premises in interactive theorem provers, and LeanDojo \cite{leandojo} and DS-Prover \cite{ds-prover} integrate models dedicated to this task.

Some of the leading models have displayed impressive prowess in proof generation, such as the HyperTree proof search \cite{HyperTree}, a new method related to the Monte Carlo tree search which uses online training to achieve the highest performance in the field.
Also, AlphaGeometry \cite{alphageometry} has recently demonstrated exceptional performance in the field of geometry proof search by combining a symbolic deduction engine, a type of computer program, with language models fine-tuned on more than 100 million geometry problems. 
We aim to preserve the overall structure of the existing proof search process, but make adaptations to the proof search algorithm to explore a wider range of proofs. Additionally, we make our adjustments to enhance performance based on ChatGPT's specific parameters, bypassing the challenges that existing algorithms face when used with ChatGPT.

\section{Background}
\subsection{Lean}

Lean is an interactive theorem prover based on dependent type theory, and allows us to use mathematical objects such as definitions, theorems, and lemmas~\cite{depend_type}. Lean utilizes tactic expressions, as do some other formal languages, which can be helpful in writing proofs. For example, \texttt{have} is a tactic that can create a new subgoal to act as a stepping stone in a theorem's proof, enabling users to construct a forward proof. Lean is popular in the mathematical community, serving as a base for active projects such as the Aesop and the Liquid Tensor experiments \cite{aesop,liquid}.
Furthermore, there is a massive library called \textit{mathlib} in Lean, which covers various mathematical topics. Considering the above aspects, we selected Lean as a proof assistant for our experiments.
Our main experiment utilizes the GPT-4 Turbo model of ChatGPT. This model generates proofs using Lean 3 syntax by default, and considering its pre-training period, it is unfamiliar with Lean 4.
Therefore, we conducted all our experiments using Lean 3, with the exception of our further analysis involving Llemma, for which we used Lean~4, following the methodology outlined in the existing Llemma paper.

\subsection{MiniF2F}
\label{subsec:MiniF2F}
The dataset miniF2F \cite{minif2f} is a cross-system benchmark for formal mathematical proof, formalized in the formal languages Lean, Metamath, and Isabelle. MiniF2F consists of problems selected from the MATH dataset~\cite{MATHdataset}; Olympiad-level problems, chosen from past AIME, AMC, and IMO questions; and CUSTOM problems, derived from topics in algebra, number theory and induction.
The MATH dataset is divided into \(5\) levels of difficulty: problems in levels \(1\) to~\(3\) require only simple calculations and straightforward approaches, while questions in levels \(4\) to \(5\) have more complex solutions. 
The Olympiad and CUSTOM problems in miniF2F range from high school-level to undergraduate-level mathematics. In general, solving the problems in these two categories necessitates more logical reasoning or stepwise proofs, sometimes by introducing premises to aid the proof. 

\subsection{Proof Search Algorithm} 
\label{subsec:Search Algorithm}

In \cite{pact,copra,leandojo}, the general proof search process of the proof-generating model via Lean is as follows:
\begin{enumerate}[label=(\arabic*)]
	\item The model takes a \textit{tactic state} which includes the hypotheses and goals in the current stage.
	\item The model generates \textit{n} many tactics to advance to the next goal or to generate new hypotheses.
	\item Each generated tactic is sent to Lean to obtain new goals or hypotheses, updating the tactic state. (We henceforth refer to this step as \textit{interaction} with Lean.)
	\item Repeat steps (1)-(3) for each new tactic state.
    \item The process is terminated when (a) goals no longer appear, indicating a valid proof has been generated, (b) all tactics fail to update the tactic state, (c) the model gives up on proving, generating \texttt{sorry}, or (d) the time limit is reached.
\end{enumerate}

Our proof search algorithm also follows the above process; the details are given in the following section.

\section{Characteristics of Our Model}
\label{subsubsec:Difference btw ours and prevs}

Often, a pre-trained model is fine-tuned on newly constructed data to create a model tailored for a specific task. However, the process of constructing a proper dataset itself is demanding. Even if the data is well-structured, fine-tuned models frequently exhibit suboptimal performance on out-of-distribution data (e.g., \cite{pact,leandojo}). 
Thus, we choose to employ ChatGPT without any additional fine-tuning, which relieves the burden of further training. Additionally, our experiments primarily utilize ChatGPT with prompt messages containing just two simple guidelines, while COPRA \cite{copra} generates the desired output through relatively complex prompt messages. 

The development of the existing models can be broadly divided into three approaches: (i) creating new training data or improving existing data \cite{pact,ds-prover,leandojo}, (ii) conducting additional training after fine-tuning \cite{HyperTree,curriculum,TacticZero}, or (iii) introducing appropriate proof search algorithms \cite{MIT,thor,DraftSketchProve,copra}. 
We choose to focus on the last method to improve performance, and as such, modify the existing algorithms to suit our purposes. We design two types of proof search algorithms following a similar process as in Section~\ref{subsec:Search Algorithm} - one with the properties of a breadth-first search, which we call the \textit{b-search}, and the other with the properties of a depth-first search, which we call the \textit{d-search}. While the b-search requires more requests to ChatGPT and more memory, it generally achieves greater performance. In contrast, the d-search usually has lower performance but requires fewer resources. We anticipate that the combination of these two types of searches with ChatGPT will lead to the exploration of a wider range of tactics. The details of our modified proof search algorithms are explained below.

The b-search follows closely with the search process outlined in Section~\ref{subsec:Search Algorithm}, with the number of generated tactics set to 64 through the ChatGPT hyperparameter ``\(n\)'' in API reference\footnote{\url{https://platform.openai.com/docs/api-reference/chat/create\#chat-create-n}}.
This method resembles the beam search-based proof search in ReProver \cite{leandojo} and Llemma \cite{llemma}.  
However, instead of looking at \(n\) many tactics through beam search as ReProver and Llemma do, the b-search generates tactics that have a very high probability of correctly continuing the proof but are likely to be redundant. This slows down the model, so we perform an additional step after step (2) in Section~\ref{subsec:Search Algorithm} to eliminate the doubles by deduplicating the generated tactics for each tactic state.
We also regulate the diversity of the generated tactics in the b-search by temperature, another ChatGPT hyperparameter. 

In the d-search, a single tactic (\(n\)=1) with the highest probability of success is generated at each step of the proof (i.e., a greedy search base). After interaction with Lean, if the tactic fails, the search returns to the initial state and reattempts a proof of the theorem. Generating all the proof steps correctly from the initial state while generating only one tactic at a time is challenging, so we alter the number of attempts (later parameterized with $k$) to generate the proof.
Although the d-search is taxing, as it involves restarting from the initial state rather than from the state that failed, we find that it can explore possible tactics more extensively through its consideration of tactics from previous steps.
Furthermore, since this method revisits a single problem multiple times, it is suitable to apply the feedback algorithm with the so-called ``Bad(O)'', as proposed in COPRA~\cite{copra}. 

Bad(O) refers to the set of tactics that are unproductive or erroneous within a specific proof state. This set ensures that, when COPRA encounters the current proof state again, it avoids generating these unsuccessful tactics. This concept helps to avoid repeating the same mistakes by recording these failed tactics, thus leading to progress in the proof generation process.
In contrast to COPRA, our d-search restarts the entire proof from the initial stage, even in cases where it fails at a later stage. Therefore, we modify the original Bad(O) feedback algorithm to complement our proof search by storing the tactic states along with their failed tactics in Bad(O), instead of just the failed tactics. Then, when the proof is restarted, the tactic state is checked against those in Bad(O) to avoid repeating past failures if a match is found. We walk through the steps for d-search with Bad(O) in a specific example in Section~\ref{sec:case_study}.

\section{Experiments}

In this section, we present the various experiments we conducted by employing ChatGPT, which is easily accessible to the general public, to achieve performance comparable to current models. 
By applying our search algorithms, we attain slightly higher performance than that of the COPRA models. Additionally, we include the results of the ReProver and COPRA models for a comprehensive comparison with our output.

\subsection{Experimental Setup}
We classify our models based on the methods used in their proof search processes, as follows:
\begin{enumerate}
	\item \textit{bChatLean}, based on b-search, and
	\item \textit{dChatLean}, based on d-search.
\end{enumerate}
We incorporate a feedback algorithm using Bad(O), as discussed in Section~\ref{subsubsec:Difference btw ours and prevs}, into the dChatLean model to get a new model, which we refer to as \textit{dChatLean+}. Additionally, we experiment with a combination of the bChatLean, dChatLean, and dChatLean+ models.

These models use ChatGPT with either GPT-4-0613 (GPT-4) or GPT-4-1106-preview (GPT-4 Turbo) as a baseline. Our implementation involves 5-shot tests, using 5 examples extracted from the miniF2F validation data. To guide ChatGPT toward proper tactic creation, we present the following prompt message:\\

\texttt{\small
	Make a proof statement in Lean3 to prove theorem using the following \\ guidelines:\\
}

\texttt{\small
- Generate only the single line of proof that immediately follows. 
}

\texttt{\small
- Do not use `sorry'.
}\\

Our models are significantly influenced by the diversity of output from ChatGPT, which is affected by the temperature setting. In our experiments, we explore two temperature values: \(0.7\) and \(1.4\). As for other hyperparameters, in bChatLean, we fix the number of tactic samplings \(n\) in a single step to 64, and in dChatLean, we investigate the effect of varying the number of attempts~\(k\) to solve one theorem, comparing results at values of \(1\), \(10\), and \(50\).

To interact with Lean, we utilize LeanDojo \cite{leandojo}, which allows parallel implementation, easily checks time limits, and is well-organized. The time limit to prove one theorem is set to \(600\) seconds, consistent with other papers. 
All experiments in this section are conducted in parallel on the CPU.
The source code for our algorithms and experiments discussed in this paper is available on GitHub\footnote{\href{https://github.com/chatlean}{https://github.com/chatlean}}.

\subsection{Main Result}
\label{our_exp}

\begin{table*}[ht]
	\caption{Pass@$k$ rates on the miniF2F test dataset}
	\label{tab:chatgpt_lean}
	\centering
	\resizebox{1.0\linewidth}{!}{
	\begin{tabular}{cccccc}
		\toprule
		Model & Baseline & \begin{tabular}{c}\# of \\Att.  (\(k\))\end{tabular} & \begin{tabular}{c}\# of \\Tactics  (\(n\))\end{tabular}  & \begin{tabular}{c}Sample\\budget\end{tabular} & Pass@$k$\\
		\midrule\midrule
        PACT \cite{pact} & similar to GPT-3 & 1 & 8 & $1 \times 8 \times 512$ & 24.6\%\\
        \midrule
        Lean Expert iteration \cite{curriculum} & similar to GPT-3 & 1 & 8 & $1 \times 8 \times 512$ & 29.6\%\\
        \midrule
		ReProver \cite{leandojo} & ByT5 & 1 & 64 & $1 \times 64 \times -$ & 26.5\%\\ 
        \midrule
        DS-Prover \cite{ds-prover} & ByT5 & 1 & $-$ & $-$ & 29.8\%\\
        \midrule
        HyperTree \cite{HyperTree} &  & 64 & 48 & $64 \times 5000$ & \bf{41.0\%}\\ 
        \midrule
        Llemma-7b \cite{llemma} & Code Llama 7b & 1 & 32 & $1 \times 32 \times 100$ & 26.23\%\\
        \midrule
		\begin{tabular}{c}COPRA\\(GPT-4 Turbo+retrieval) \cite{copra}\end{tabular} & GPT-4 T(urbo) & 1 & 1 & $1 \times 60$ & 26.64\%\\
        \midrule
		\begin{tabular}{c}COPRA+One-Shot\\+Informal-One-Shot\\(GPT-4 Turbo+retrieval\\+informal proof) \cite{copra}\end{tabular} & GPT-4 T & 1 & 1 & $1 \times 100$ & 29.92\%\\
		\midrule\midrule
        bChatLean & GPT-4 T & 1 & 64 & $1 \times 64 \times -$ & 29.10\%\\
		\midrule
        dChatLean & GPT-4 T & 50 & 1 & $50 \times 1 \times -$ & 23.77\%\\
        \midrule
        dChatLean+ & GPT-4 T & 50 & 1 & $50 \times 1 \times -$ & 25.00\%\\
		\midrule
		bChatLean \& dChatLean+& \bf{GPT-4 T} & \bf{1 \& 50} &  \bf{64 \& 1} & \begin{tabular}{c} (\bf{1} $\times$ \bf{64} $\times -$) \\ \bf{+} (\bf{50} $\times$ \bf{1} $\times -$)\end{tabular} 
        & \bf{31.15\%}\\
		\bottomrule
	\end{tabular}
}
\end{table*}

Our results are presented in Table \ref{tab:chatgpt_lean}. The results of our models shown in this table are all obtained using GPT-4 Turbo with a temperature setting of \(T=1.4\). We display the \textit{pass@$k$ rate} of each model as presented in each paper for the miniF2F Lean test dataset. The pass@$k$ rate, the most widely used metric for evaluating proof-generating models, represents the proportion of completed proofs over the entire dataset, where \(k\) denotes the number of attempts to generate a proof made on each problem. We employ specific terminology in our table for simplification.
\begin{itemize}
	\vspace{0.2\baselineskip}
	\vspace{0.2\baselineskip}
    \item[-] \textit{\# of Att. (\(k\))}: The number of attempts \(k\) to prove a problem, the same \(k\) as in a pass@$k$ rate.
    \vspace{0.2\baselineskip}
	\item[-] \textit{\# of Tactics (\(n\))}: The number of tactic samples \(n\) a model generates upon receiving one tactic state. It is similar in notion to the number of beams in a typical beam search.
	\vspace{0.2\baselineskip}
    \item[-] \textit{Sample budget}: A measure of computational complexity, generally in the form of $k \times n \times e$, where $e$ denotes the number of calls to the baseline model. Alternative forms and a more detailed explanation are provided below.
\end{itemize}

Current proof models use a variety of proof generation algorithms, making it challenging to conduct a fair comparison of their computational complexities. We thus compute the ``sample budget'' based on \cite{deepseek-prover}.
The general form for the sample budget is $k \times n \times e$, but in the case of HyperTree and COPRA, the sample budget is represented as $k \times e$. Also, in our table, ``--'' represents a dynamic value that varies depending on the situation and has no fixed upper limit. 

Although our models use a dynamic value for the parameter $e$, as a reference, if we retrospectively examine the experimental results and calculate the average number of calls, this value can be compared to the $e$ value of other models. To give a few of the values, the average number of calls was $1.664$ for bChatLean and $1.057$ for dChatLean+. 

Our dChatLean models show lower performance than the bChatLean model. In particular, as can be seen from Table \ref{tab:chatgpt_lean}, the best performance for dChatLean is lower than the 29.10\% pass rate achieved by bChatLean, reaching a peak pass rate of 25.00\% through its natural synergy with Bad(O) in dChatLean+. 
The lower performance from dChatLean can be attributed to the relative difficulty of the dChatLean proof search problem, as it requires restarting from the beginning if an intermediate step fails. 

Since the two proof searches described above use separate strategies to find proofs, we expect that combining the two models will open up room for a wider range of tactics to explore. Thus, we couple bChatLean with dChatLean+, both under GPT-4 Turbo (\(k=1, n=64\) for bChatLean and \(k=50, n=1\) for dChatLean+), yielding an impressive performance of 31.15\%. This pass rate surpasses all values in Table \ref{tab:chatgpt_lean} except for in the case of the HyperTree model, which uses a much larger sample budget. 
Out of the 244 theorems attempted, 56 were successfully solved by both of our models, 15 were only solved in bChatLean, and 5 were only solved in dChatLean+. Furthermore, out of the proven theorems, 21 theorems have not been proved in the recent miniF2F GitHub 
repository\footnote{\href{https://github.com/facebookresearch/miniF2F/tree/88b0189c213f3bd4b020c215f5bf0d847ee8d7a0}{GitHub repository for MiniF2F dataset}}.

Despite the high performance ultimately obtained, there are still some limitations in our models. To name one in particular, while a simplified prompt is handy for our experiment, it occasionally fails to deliver our intention to the model. For example, we introduce a guideline of ``\texttt{Generate only the single line of proof that immediately follows}'' to facilitate step-by-step theorem proving. Contrary to our purposes, the proof for the solved problems often manifested as a single line of output, achieved by combining various tactics through the generous use of semicolons. Consequently, we think that our models can report a higher pass rate if slightly more detailed instructions are given. Many other areas can be improved, but we leave these to future works.

\subsection{Ablation Studies}

\subsubsection*{On the number of attempts in dChatLean}

Generating proofs by using dChatLean is significantly more challenging than with other models that utilize breadth-first search or depth-first search algorithms. One of the reasons for this is that dChatLean restarts the proof generation from the initial state if it fails. Additionally, any minor typos in dChatLean's tactic generation can critically impact the model's pass rate. We observe that, as the number of attempts $k$ made on a problem increases, dChatLean's proof generation stabilizes, directly enhancing the overall performance of the model (see Table~\ref{tab:ablation_k}). In GPT-4 Turbo with temperature $T=0.7$, an increase of $k$ from 10 to 50 yields a performance increase of 6.97\%. The same trend holds for GPT-4 Turbo under $T=1.4$, as the addition of attempts from 10 to 50 shows a performance rise of 8.61\%. This notable increase can be attributed to the model attempting many proof directions by generating more diverse tactics multiple times, confirming that increasing the number of attempts on a proof produces a higher pass rate. 

\begin{table}[ht]
	\caption{Ablation study for the number of attempts}
	\label{tab:ablation_k}
	\centering
	\resizebox{0.6\linewidth}{!}{
	\begin{tabular}{c|ccc|c}
		\toprule
		Model & Baseline & Temp. (\(T\)) & \# of Att.  (\(k\))  & Pass Rate\\
		\midrule\midrule
		\multirow{4}{*}{dChatLean}& \multirow{2}{*}{GPT-4 T} & \multirow{2}{*}{0.7} & 10 & 13.93\%\\
		& & & 50 & 20.90\%\\
            \cline{2-5}
		& \multirow{2}{*}{GPT-4 T} & \multirow{2}{*}{1.4} & 10 & 15.16\%\\
		& & & 50 & 23.77\%\\
		\bottomrule
	\end{tabular}
 }
\end{table}

\subsubsection*{On tactic diversity}

Since dChatLean only produces one tactic at a time, it is important to explore multiple possibilities by generating a variety of tactics in subsequent attempts. Additionally, since bChatLean uses ChatGPT's $n$ hyperparameter, it is necessary for the model to produce more diverse outputs to avoid repeated attempts. In this context, temperature is a crucial parameter that influences the model's pass rate.

As shown in Table~\ref{tab:ablation_t}, in our dChatLean model, GPT-4 with $k=10$ and GPT-4 Turbo with $k=10$ both reflect a performance gain of about 1.23\% when $T$ is raised from 0.7 to 1.4. GPT-4 Turbo with $k=50$ when $T=1.4$ yielded a performance increase of about 2.87\% from its performance when $T=0.7$. 
Hence, we see that the influence of temperature changes on the pass rate is more substantial when the number of attempts is higher, highlighting the benefits of a wider range of attempts on performance.
This tendency of achieving better performance with higher temperature is mirrored in our bChatLean model, as GPT-4 Turbo with $T$ fixed at 1.4 reflects a pass rate 1.23\% higher than when $T$ is 0.7.

\begin{table}[ht]
\centering
 	\caption{Ablation study for the temperature}
	\label{tab:ablation_t}
	\resizebox{0.6\linewidth}{!}{
	\begin{tabular}{c|lcc|c}
		\toprule
		Model & Baseline & \# of Att.  (\(k\)) & Temp. (\(T\)) & Pass Rate\\
		\midrule\midrule
		\multirow{2}{*}{bChatLean}& \multirow{2}{*}{GPT-4 T} & \multirow{2}{*}{1} & 0.7 & 27.87\%\\
		&  &  & 1.4 & 29.10\%\\
		\midrule
		\multirow{6}{*}{dChatLean}& \multirow{2}{*}{GPT-4} & \multirow{2}{*}{10} & 0.7 & 14.75\%\\
		&  & & 1.4 & 15.98\%\\
            \cline{2-5}
		& \multirow{2}{*}{GPT-4 T} & \multirow{2}{*}{10}& 0.7 & 13.93\%\\
        &  &  & 1.4 & 15.16\%\\
            \cline{2-5}
		& \multirow{2}{*}{GPT-4 T} & \multirow{2}{*}{50} & 0.7 & 20.90\%\\
		& &   & 1.4 &23.77\%\\
		\bottomrule
	\end{tabular}
	}
\end{table}

\begin{table}
\centering
	\caption{Ablation for the feedback algorithm by Bad(O)}
	\label{tab:ablation_bado}
	\resizebox{0.6\linewidth}{!}{
	\begin{tabular}{c|ccc|c}
		\toprule
		Model & Baseline & Temp. (\(T\)) & \# of Att.  (\(k\))  & Pass Rate\\
		\midrule\midrule
            dChatLean & \multirow{2}{*}{GPT-4 T} & \multirow{2}{*}{1.4} & \multirow{2}{*}{10} & 15.16\%\\
		dChatLean+ & & & & 18.85\%\\
            \cline{1-5}
		dChatLean & \multirow{2}{*}{GPT-4 T} & \multirow{2}{*}{1.4} & \multirow{2}{*}{50} & 23.77\%\\
		dChatLean+ &  &  &  & 25.00\%\\
		\bottomrule
	\end{tabular}
	}
\end{table}

\subsubsection*{On the feedback algorithm in dChatLean}

Due to dChatLean's characteristic of returning to the initial state to reattempt the proof, it is natural that some tactic states reappear during the proof search. Thus, the feedback algorithm by Bad(O) plays a vital role in increasing the pass rate of the model by avoiding previously failed tactics and instead coaxing the model to consider new tactics.
Table~\ref{tab:ablation_bado} shows the comparison between pass rates for dChatLean, which does not use a feedback algorithm, and dChatLean+, which utilizes the feedback algorithm with Bad(O). 
When the number of attempts $k$ is 10, the inclusion of the feedback algorithm helps dChatLean+ achieve a pass rate 3.69\% higher than that of dChatLean. Setting $k$ to 50, dChatLean+ records a performance 1.23\% above dChatLean's.
In Section~\ref{sec:case_study}, we provide a specific example to illustrate the process of generating a proof using Bad(O) in dChatLean+ (c.f. Figure~\ref{fig:case_study_2_1} and Figure~\ref{fig:case_study_2_2}).

\subsection{Evaluation Across Datasets and LLMs}
\label{Evaluation}

We evaluate our models on the ProofNet dataset and the 2023 AMC 12 problems to verify our models' abilities to generate proofs for problems outside of the scope covered by miniF2F. We also assess our algorithms using another language model, Llemma, in place of ChatGPT. Despite the simplicity of our models, the results demonstrate performance comparable to established benchmarks, effectively addressing mathematical problems across diverse contexts and settings.

\subsubsection*{On ProofNet}
We used our bChatLean and dChatLean+ coupled model on ProofNet\footnote{\href{https://github.com/zhangir-azerbayev/ProofNet/tree/02b47219879da3a4fcb4d6ad5f2f29f384cf2de0}{GitHub repository for ProofNet dataset}}, a benchmark for undergraduate-level mathematics, covering Analysis, Abstract Algebra, Linear Algebra, and Topology, as well as a topic labeled ``Examinations'' that consists of some Putnam Competition problems. The ProofNet set is composed of 350 problems drawn from mathematics textbooks and formalized in Lean 3. When running our model on this benchmark, we achieve an overall performance of 13.14\%, as seen in Table \ref{tab:ProofNet_results}, and generate 35 new Lean proofs.
The performance in Table \ref{tab:ProofNet_results} outlines the pass rate across different mathematical topics. 
In particular, our model performs best with Analysis problems, with a pass rate of 15.91\%, and conversely, finds Linear Algebra and Topology problems more challenging, with pass rates of 10.00\% and 10.71\%, respectively. This domain-based comparison informs us of the specific strengths and weaknesses within our models, providing direction for future enhancements.

\begin{table}[ht]
\centering
    \caption{Pass rates on ProofNet by topic}
    \label{tab:ProofNet_results}
	\resizebox{0.5\linewidth}{!}{
    \begin{tabular}{c|c|c|c}
        \toprule
        Topic & ~Theorems~ & ~Proved~ & ~Pass rate \\
        \midrule\midrule
        Analysis & 88 & 14 & 15.91\% \\
        ~Abstract Algebra~ & 162 & 21 & 12.96\% \\
        Linear Algebra & 28 & 3 & 10.71\% \\
        Topology & 60 & 6 & 10.00\% \\
        Examinations & 12 & 2 & 16.67\% \\
        \midrule
        Total & 350 & 46 & 13.14\% \\
        \bottomrule
    \end{tabular}
    }
\end{table}

\subsubsection*{On AMC 12}
To conduct our evaluation, we selected problems from the 2023 AMC 12A\footnote{\href{https://artofproblemsolving.com/wiki/index.php/2023_AMC_12A}{2023 AMC 12A Problems and Solutions}} and 12B\footnote{\href{https://artofproblemsolving.com/wiki/index.php/2023_AMC_12B}{2023 AMC 12B Problems and Solutions}} exams, both of which make up the high school mathematics problem-solving competition that acts as a first step towards Olympiad-level tests. 
These problems were published after the training for the ChatGPT version that we use for this paper, and thus were outside of the possible scope of problems that could have slipped into the training set for the model.
We formalized 26 problems from this competition into Lean 3 using ChatGPT, and manually verified that the formalized problems were free of errors and retained their original meanings\footnote{\href{https://github.com/chatlean/AMC12\_2023\_Lean}{https://github.com/chatlean/AMC12\_2023\_Lean}}. After running the bChatLean \& dChatLean+ coupled model on this dataset, we obtained 6 successful proofs, giving us a pass rate of 23.07\%. 
Although we do not have access to the data that ChatGPT was trained on, we test our models on some newly published problems and note that we achieve a comparable pass rate as on the miniF2F dataset. This demonstrates our models' applicability to new problems, highlighting the potential that these models can later have in real-world applications. 

\subsubsection*{Using Llemma} 
Though our models are based on search algorithms that specifically cater to ChatGPT and its properties, we verify our proof searching algorithms' capabilities with the mathematics-specialized large language model Llemma \cite{llemma}. In parallel with the ChatGPT-based models, we call the Llemma-based models bLlemLean, dLlemLean, and dLlemLean+. These models are based on the b-search, d-search, and the d-search with Bad(O), respectively. They continue to use Lean as their formal language, but since the established performance of Llemma is dependent on Lean~4, we adopt Lean~4 in this experiment rather than Lean~3. The miniF2F pass rates for these models and their combinations are outlined in Table \ref{tab:llemma_results}.

Unlike with our ChatGPT models, our dLlemLean+ model performs worse than our dLlemLean model, indicating that the d-search method suffers with the introduction of Bad(O). We can attribute this to the fact that Bad(O) generally makes longer prompts, which Llemma struggles with due to the far fewer parameters used in comparison to ChatGPT. 
However, Bad(O) has its advantages when d-search is used together with b-search, as the bLlemLean and dLlemLean combined model performs a little worse than the bLlemLean and dLlemLean+ combined model does. This suggests that Bad(O) still aids d-search in finding proofs that the b-search fails to find alone.

We achieved the highest performance of 28.28\% using the combination of bLlemLean and dLlemLean+, surpassing the benchmark of 26.23\% previously reported for Llemma. Although our proof search algorithms were specially customized to enhance the ChatGPT output for our purposes, that the Llemma-based models perform comparatively well conveys that our algorithms are robust to language model substitutions.

\begin{table}
\centering
    \caption{MiniF2F test pass rate on Llemma}
    \label{tab:llemma_results}
	\resizebox{0.45\linewidth}{!}{
	\begin{tabular}{c|c}
		\toprule
        Model & ~Pass Rate \\
        \midrule\midrule
        Llemma \cite{llemma} & 26.23\% \\
        \midrule
        bLlemLean & 26.64\% \\ 
        dLlemLean & 22.54\% \\
        dLlemLean+ & 21.31\% \\
        bLlemLean \& dLlemLean & 27.46\% \\
        bLlemLean \& dLlemLean+ & \textbf{28.28\%} \\ 
        \bottomrule
	\end{tabular}
	}
\end{table}

\section{Additional Analysis with Examples}
\label{sec:case_study}

In this section, we analyze our results based on specific criteria with examples.

\subsection*{Considering the number of attempts in dChatLean+}

\begin{figure}[ht]
	\centering
	\includegraphics[width=0.35\linewidth]{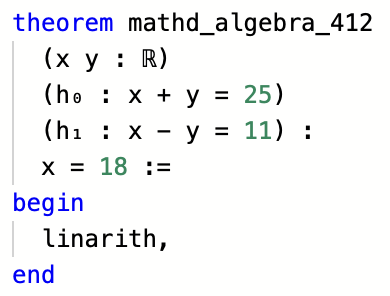}
    \includegraphics[width=0.55\linewidth]{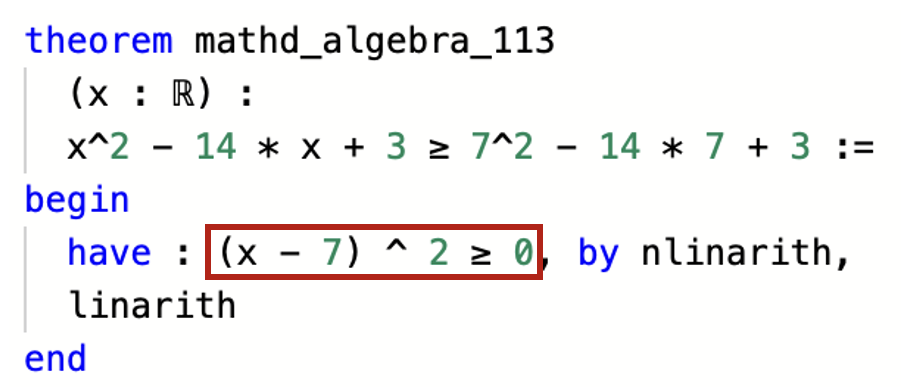}
	\caption{Examples considering numbers of attempts taken by dChatLean+}
	\label{fig:case_study_1}
\end{figure}

In our experiments with dChatLean+, where $k$ was set to be strictly greater than~1, we found that for the problems solved in only one attempt, the proofs involved either performing a simple calculation or simply solving a system of equations, as illustrated in the left example of Figure~\ref{fig:case_study_1}. In contrast, problems solved with more than one attempt included proofs demonstrating that given properties hold, as shown in the right example of Figure~\ref{fig:case_study_1}. 
These findings suggest that increasing the number of attempts in dChatLean+ may enable the model to identify approaches that require additional premises beyond basic calculations.

Moreover, we observe that, for the problems with successful proofs, the average number of attempts taken with dChatLean is 10.3, while just 9.3 with dChatLean+. (These values are based on our observations but are not explicitly presented in the paper.) This indicates that the feedback algorithm using Bad(O) guides the model to produce fewer failures when searching for a solution.
Table~\ref{tab:trial_numbers} shows the number of attempts taken by both models to solve a few problems for which they produce similar proofs.

\begin{table}[ht]
	\caption{Number of attempts taken by dChatLean and dChatLean+}
	\label{tab:trial_numbers}
	\centering
	\resizebox{0.7\linewidth}{!}{
	\begin{tabular}{c|c|c}
		\toprule
        Name of problem & dChatLean & dChatLean+\\
		\midrule\midrule
		mathd$\_$algebra$\_$478 & 5 & 2 \\
        algebra$\_$sqineq$\_$unitcircatbpabsamblt1 & 12 & 2\\
		mathd$\_$algebra$\_$113 & 47 & 17\\
		\bottomrule
	\end{tabular}}
\end{table}

\subsection*{Considering the feedback algorithm}

\begin{figure*}[ht]
    \centering
    \includegraphics[width=\linewidth]{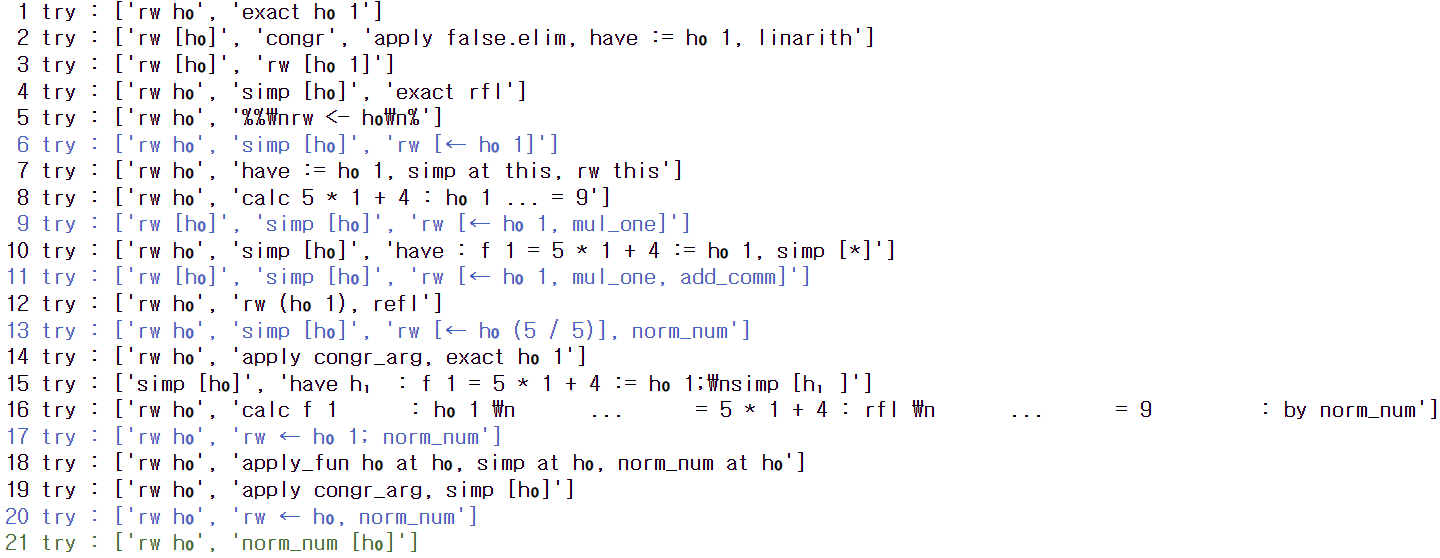}
    \caption{The process of generating a proof for ``mathd\_algebra\_171'' using the Bad(O)-based feedback algorithm. The highlighted lines indicate the trials selected to be further illustrated in Figure~\ref{fig:case_study_2_2}. The final proof is given in green, while other selected trials are in blue.}
    \label{fig:case_study_2_1}
\end{figure*}

\begin{figure*}[ht]
    \centering
    \includegraphics[width=\linewidth]{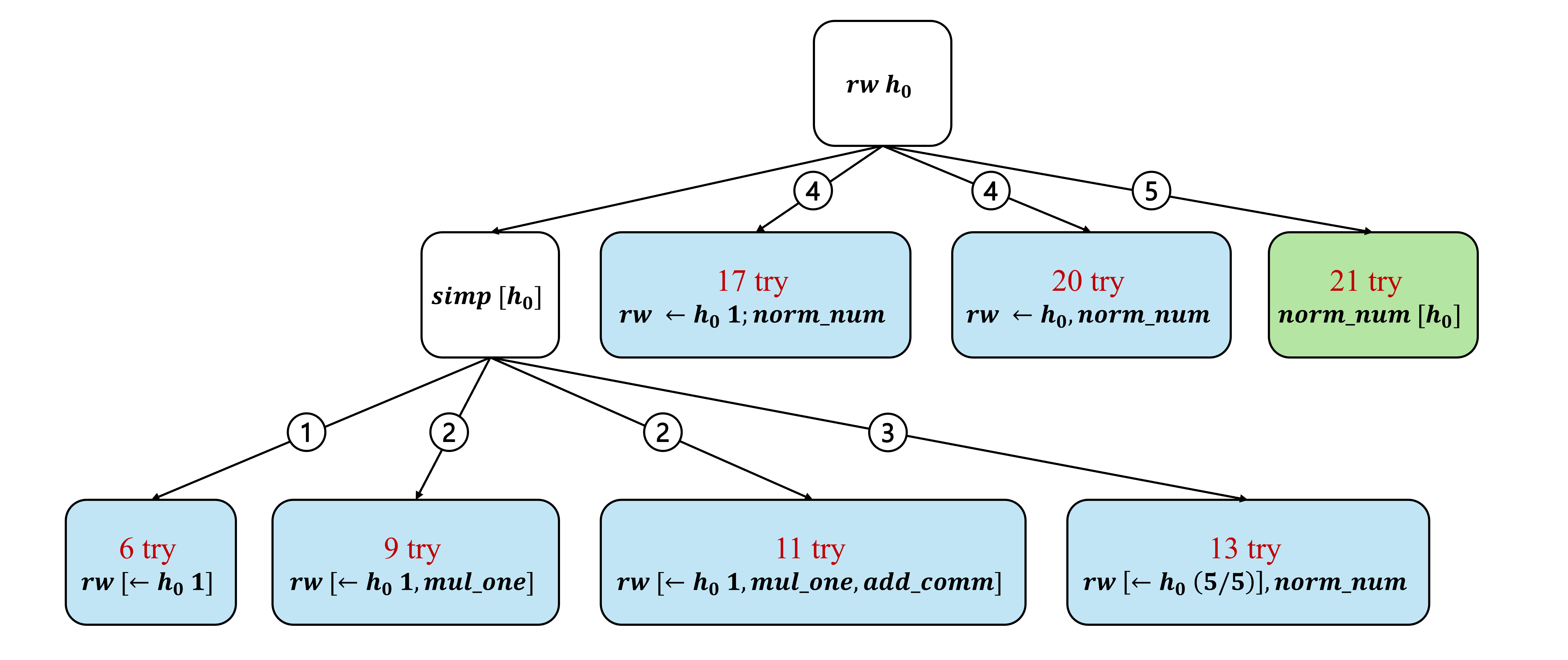}
    \caption{Selected proof trials highlighted in blue, also displayed in Figure~\ref{fig:case_study_2_1}. The final proof is highlighted in green.}
    \label{fig:case_study_2_2}
\end{figure*}

The proof generation process using the Bad(O)-based feedback algorithm in dChatLean+ can be seen in Figure~\ref{fig:case_study_2_1}, where the model completed the proof for the miniF2F problem ``mathd\_algebra\_171'' in 21 attempts.
The completed proof, 21 try, begins with \texttt{rw h\textsubscript{0}}, and one can note many previous failed attempts that also started with \texttt{rw h\textsubscript{0}}. These attempts were stored in Bad(O) to avoid repeating the same failure when faced with the same tactic state. 
We depict a few selected trials in Figure~\ref{fig:case_study_2_2} and walk through the proof search process below.

\begin{enumerate}[label=\small\protect\textcircled{\small\arabic*}]
\item On the sixth trial, the model reached \texttt{rw h\textsubscript{0}} and \texttt{simp [h\textsubscript{0}]}. To continue the proof, it then generated the tactic \texttt{rw [\textnormal{\textleftarrow} h\textsubscript{0} 1]}, but resulted in failure. This tactic \texttt{rw [\textnormal{\textleftarrow} h\textsubscript{0} 1]} was thus stored in Bad(O) for the model to refer to in any subsequent attempts to solve the same tactic state--or goal--that \texttt{rw [\textnormal{\textleftarrow} h\textsubscript{0} 1]} was brought in to address.
\vspace{0.2\baselineskip}
\item The same tactic state appeared in trials 9 and 11, and the model avoided the tactics in Bad(O) to try new tactics, generating \texttt{rw [\textnormal{\textleftarrow} h\textsubscript{0} 1, mul\_one]} for trial 9 and \texttt{rw [\textnormal{\textleftarrow} h\textsubscript{0} 1, mul\_one, add\_comm]} for trial 11. However, these also resulted in failures, and these new tactics were added to Bad(O).
\vspace{0.2\baselineskip}
\item After this series of failures, the model decided to step out and seek a solution for the tactic state using \texttt{norm\_num}, as seen in trial 13 with \texttt{rw [\textnormal{\textleftarrow} h\textsubscript{0} (5/5)], norm\_num}. This also resulted in failure, adding this last tactic to Bad(O) as well.
\item The model decided to then step out of the tactic state involving \texttt{simp [h\textsubscript{0}]} and instead work in the tactic state resulting from \texttt{rw h\textsubscript{0}}. Trials 17 and 20 generated tactics using \texttt{norm\_num}, but resulted in failures, which were stored in Bad(O) under this new tactic state. 
\vspace{0.2\baselineskip}
\item Finally, under guidance from Bad(O), the model produced a successful proof in trial 21. After seeing that all goals have been cleared, the proof search algorithm terminates and returns the successful proof.
\end{enumerate}

As demonstrated in this example, the feedback algorithm in dChatLean+ allows the model to adapt failed tactics or change the direction of the proof by referring to previous attempts.

\subsection*{Considering challenging problems}

\begin{figure}[ht]
    \centering
    \includegraphics[width=0.4\linewidth]{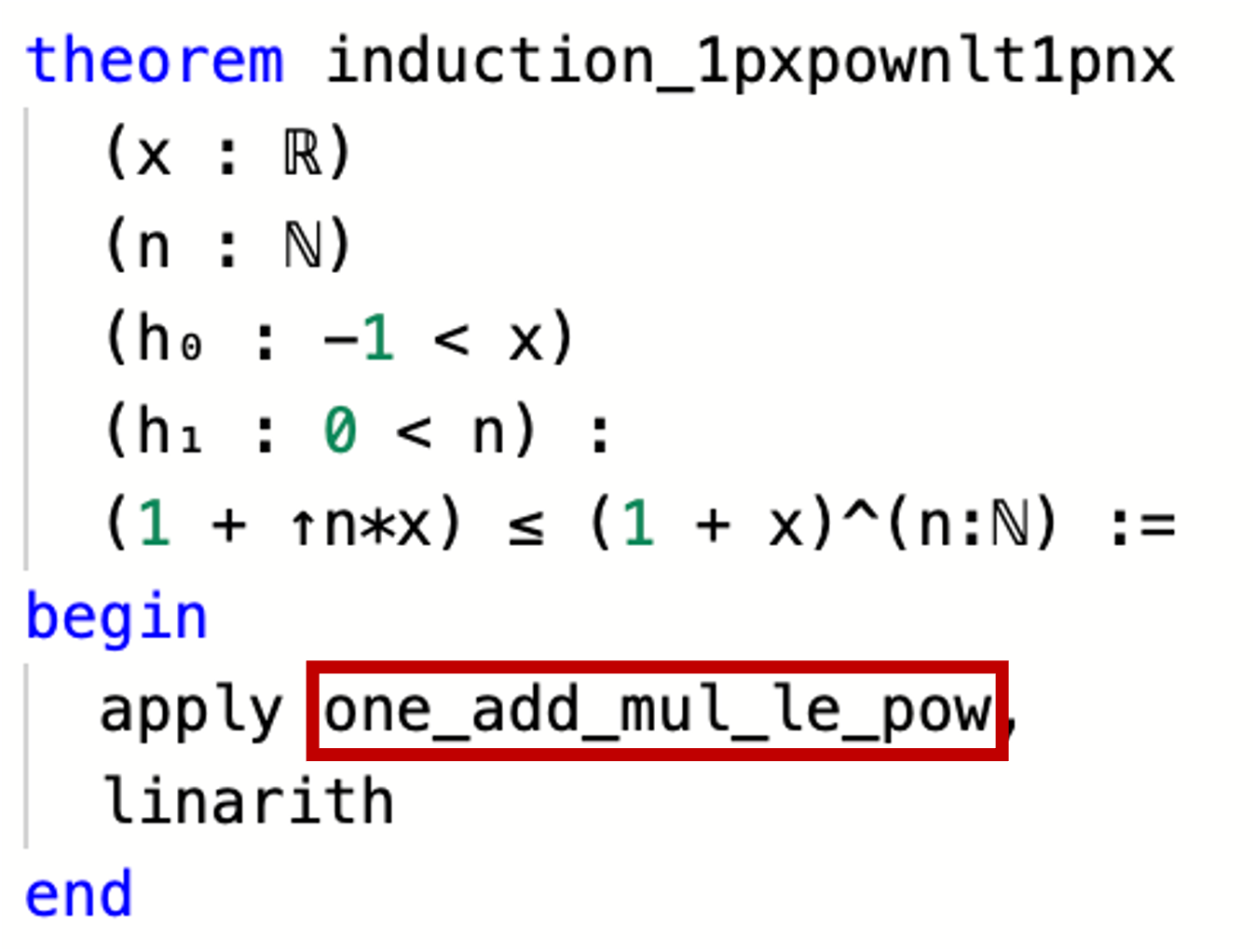}
    \caption{The proof generated by bChatLean for Bernoulli's inequality}
    \label{fig:case_study_bchatlean}
\end{figure}

Our bChatLean model solves a problem named ``induction\_1pxpownlt1pnx'' from the miniF2F dataset with the proof presented in Figure~\ref{fig:case_study_bchatlean}. 
This problem corresponds to Bernoulli's inequality, which we believe is intended to be proved using mathematical induction. However, the model uses the pre-defined theorem 
\texttt{one\_add\_mul\_le\_pow},
which corresponds to Bernoulli's inequality in the mathlib library. 
While the problem was technically solved, the resulting proof may be considered unsatisfactory.
The outcome suggests that the model may have leveraged prior exposure to the library during training, leading it to apply the theorem directly. Conversely, we found that the ReProver model, which is probably less influenced by such prior exposure due to its smaller size, cannot prove this problem. This raises several points for consideration: there is still a lack of algorithms for small-scale models that address problems of this level, and even if a problem can be solved when scaled up in size, it is important to consider whether the solution aligns with the intended solution. Additionally, mitigating the influence of pre-existing knowledge in large-scale models will be crucial.

\section{Conclusion}

In this paper, we proposed two simple proof search methods: the b-search and the d-search. 
These gave rise to our bChatLean and dChatLean models, which employ a pre-trained large language model in ChatGPT and achieve comparable performance to that of state-of-the-art models even in the absence of fine-tuning, as demonstrated on the miniF2F dataset. 
In addition, we improved the performance of these two models by coupling bChatLean with dChatLean+, which introduces a feedback algorithm into dChatLean. 
To test the capabilities of our approach, we applied our models and proof search algorithms on other datasets, namely ProofNet and the 2023 AMC 12, and to the other large language model, Llemma.
We analyzed our models through ablation studies on factors such as the number of attempts, the temperature, and our feedback algorithm. Moreover, we selected examples to illustrate and scrutinize the models' results and presented further consideration for future research in the field of formal mathematical proof.

\section{Acknowledgments}
T. Hur--This work was partly supported by the Institute of Information \& communications Technology Planning \& Evaluation (IITP) grant funded by the Korea government (MSIT) (No. RS-2023-00221236, Development of document understanding and information extraction model using multi-modal artificial intelligence technology).

Y. Hur--This work was partly supported by the National Research Foundation of Korea (NRF) (Grant Number 2021R1A2C1007598), by the National Supercomputing Center with supercomputing resources including technical support (Grant Number: KSC-2023-CRE-0087), and by the Visiting Professorship at Korea Institute for Advanced Study.

H. Lim--This work was partly supported by the Hausdorff Research Institute for Mathematics (HIM) funded by the Deutsche Forschungsgemeinschaft (DFG, German Research Foundation) under Germany's Excellence Strategy -- EXC-2047/1 -- 390685813.

%

\end{document}